\documentclass[onecolumn]{emulateapj}
\usepackage{longtable}
\usepackage{graphics}
\usepackage{rotating}
\usepackage{amsmath}
\usepackage{txfonts}
\usepackage{natbib}
\usepackage{hyperref}
\def\gtorder{\mathrel{\raise.3ex\hbox{$>$}\mkern-14mu
             \lower0.6ex\hbox{$\sim$}}}
\def\ltorder{\mathrel{\raise.3ex\hbox{$<$}\mkern-14mu
             \lower0.6ex\hbox{$\sim$}}}

\slugcomment{\today}

\shorttitle{ngVLA memo \#28}

\shortauthors{ngVLA memo \#28}

\begin{document}

\title{\lowercase{ng}VLA memo \#28 \\Host galaxies and relativistic ejecta of compact binary mergers in the \lowercase{ng}VLA era}
\author{Alessandra~Corsi\altaffilmark{1}, Dale~A.~Frail\altaffilmark{2}, Benjamin~J.~Owen\altaffilmark{1}, David~J.~Sand\altaffilmark{1,3}, Richard~O'Shaughnessy\altaffilmark{4}, Eric~J.~Murphy\altaffilmark{5}}
 \altaffiltext{1}{Department of Physics and Astronomy, Texas Tech University, Box 1051, Lubbock, TX 79409-1051, USA; e-mail: alessandra.corsi@ttu.edu}
\altaffiltext{2}{National Radio Astronomy Observatory, P.O. Box O, Socorro, NM 87801, USA.}
\altaffiltext{3}{Department of Astronomy and Steward Observatory, University of Arizona, 933 N Cherry Ave, Tucson, AZ 85719, USA.}
\altaffiltext{4}{Center  for  Computational  Relativity  and  Gravitation, Rochester  Institute  of  Technology,  Rochester,  NY  14623,  USA.}
\altaffiltext{5}{National Radio Astronomy Observatory, 520 Edgemont Rd, Charlottesville, VA 22903, USA.}

\begin{abstract}
We present the results of a community study aimed at exploring some of the scientific opportunities that the next generation Very Large Array (ngVLA) could open in the field of multi-messenger time-domain astronomy. We focus on compact binary mergers, golden astrophysical targets of ground-based gravitational wave (GW) detectors such as advanced LIGO.  A decade from now, a large number of these mergers is likely to be discovered by a world-wide network of GW detectors. This will enable the identification of host galaxies, either directly through detection of electromagnetic (EM) counterparts, or indirectly by probing potential anisotropies in the spatial distribution of mergers. Identifying the host galaxy population of GW mergers would provide a way to constrain the efficiency of various binary neutron star (NS) or black hole (BH) formation scenarios, and the merger timescale distributions as linked to merger rates in early- and late-type galaxies.  We discuss how a radio array with $\approx 10\times$ the sensitivity of the current Karl G. Jansky VLA and $\approx 10\times$ the resolution, would enable resolved radio continuum studies of binary merger hosts, probing regions of the galaxy undergoing star formation (which can be heavily obscured by dust and gas), AGN components, and mapping the offset distribution of the mergers with respect to the host galaxy light. For compact binary mergers containing at least one NS and with associated EM counterparts, we show how the ngVLA would enable direct size measurements of the relativistic merger ejecta and probe, for the first time directly, their dynamics.
\end{abstract}

\section{Introduction}
The results summarized in this report
follow a solicitation from the National Radio Astronomy Observatory to develop community-based studies for a future U.S.-led radio
telescope, the ngVLA\footnote{https://science.nrao.edu/futures/ngvla}. Building on the scientific and technical legacy of the Jansky VLA\footnote{https://science.nrao.edu/facilities/vla} and of the Atacama Large Millimeter Array\footnote{http://www.almaobservatory.org/en/home/} (ALMA),  the ngVLA will have $\approx 10\times$ the collecting area of the
Jansky VLA, operate at frequencies from 1\,GHz (30\,cm) to 116\,GHz (2.6\,mm) with up to 20 GHz of bandwidth, possess a compact core for high
surface-brightness sensitivity, and extended baselines of at least hundreds of kilometers and ultimately across the continent to
provide high-resolution imaging \citep{BolattoMemo}.  As underlined by \citet{BowerMemo}, the ngVLA could bring transformational results by enabling the exploration of the dynamic radio sky with unprecedented sensitivity and resolution. Here we discuss two new scientific opportunities that would emerge in time-domain astrophysics if a facility like the ngVLA were to work in tandem with ground-based GW detectors:
\begin{enumerate}
\item Unraveling the physics behind the progenitors of BH-BH, NS-NS, and BH-NS mergers via host galaxy studies at radio wavelengths;
\item Enabling direct size measurements (and thus direct dynamics constraints) of relativistic ejecta from NS-NS and/or BH-NS mergers (with ngVLA plus VLBI stations).
\end{enumerate}
In what follows, we briefly describe the scientific landscape expected to be realized when the ngVLA may become operational (Sec.\,\ref{sec:1}). Then, we discuss radio studies of host galaxies of binary mergers (Sec.\,\ref{sec:2}), and of relativistic ejecta (Sec.\,\ref{sec:3}).  Finally, in Sec.\,\ref{sec:4}, we summarize and conclude. 

We note that while this report was being submitted, the first NS-NS merger was discovered by the advanced LIGO and Virgo detectors, and found to be associated with an EM counterpart detected at all wavelengths \citep[e.g.,][and references therein]{GRB,GW170817Discovery,MMApaper}. We have thus updated our discussion to briefly refer to GW170817 where relevant in the context of this study.

\section{The post-2027 scientific landscape}
\label{sec:1}
Our discussion of compact binary mergers and their relativistic ejecta in the ngVLA era is framed within the scientific landscape of a decade or more in the future. In light of the recent BBH detections by the advanced Laser Interferometer Gravitational-wave Observatory\footnote{www.ligo.org} (LIGO) and Virgo \citep[][]{GW151226,GW150914,GW170104,GW170814}, and of the first multi-messenger detection of a NS-NS merger \citep{GRB,GW170817Discovery,MMApaper} we expect that about 10 years from now the field of time-domain astronomy will have fully transitioned to time-domain GW astrophysics.  Indeed, it is likely that a decade from now the GW window will be routinely enabling multi-messenger studies of the transient sky. The network of ground-based GW detectors will include Virgo\footnote{https://www.ego-gw.it/public/about/whatIs.aspx} operating at nominal advanced sensitivity, the two advanced LIGO detectors likely in their so-called plus configuration\footnote{https://dcc.ligo.org/LIGO-T1700231/public} (which foresees a factor of $\sim 5$ increase in event rates with respect to advanced LIGO at full sensitivity), the Kamioka Gravitational Wave Detector\footnote{http://gwcenter.icrr.u-tokyo.ac.jp/en/} (KAGRA), and LIGO India\footnote{http://www.gw-indigo.org/tiki-index.php?page=LIGO-India}. This world-wide network of detectors will be identifying potentially tens to hundreds of GW in-spirals and mergers in the local universe, with localization areas of order $\lesssim 10$\,deg$^2$ \citep[a factor of $\gtrsim 10$ better than today;][]{LivRev}. 

Our understanding of the transient sky will also have been informed and shaped by the findings of several (upcoming or planned) synoptic optical surveys, such as e.g., the Zwicky Transient Facility\footnote{https://www.ptf.caltech.edu/ztf} and the Large Synoptic Survey Telescope\footnote{https://www.lsst.org/} (LSST). Particularly, when starting full operations in circa 2023 and for the following 10 years, the LSST will  discover tens of  thousands  of  astrophysical  transients every night down  to  a  limiting  magnitude of $r\approx  24.7$\,mag \citep{LSST}. Thus, by the time the ngVLA will start operations, we expect to have a better census of the optical transient sky, including statistically significant samples ($\gtrsim 10$ objects) of rare and faint transients.  Among other major EM facilities likely to be contributing observations at various wavelengths in the ngVLA era are ALMA, which will be observing in the mm, the Square Kilometer Array\footnote{http://skatelescope.org/} \citep[SKA; e.g., ][]{SKA}, and the Wide Field Infrared Survey Telescope\footnote{https://wfirst.gsfc.nasa.gov/} (WFIRST) operating in the NIR-IR. The James Web Space Telescope\footnote{https://www.jwst.nasa.gov/} (JWST) could also still (after its nominal mission duration) be providing sub-arsecond resolution and unprecedented sensitivity from long-wavelength visible light to near/mid-infrared. 

\begin{figure}
\begin{center}
\includegraphics[width=13cm]{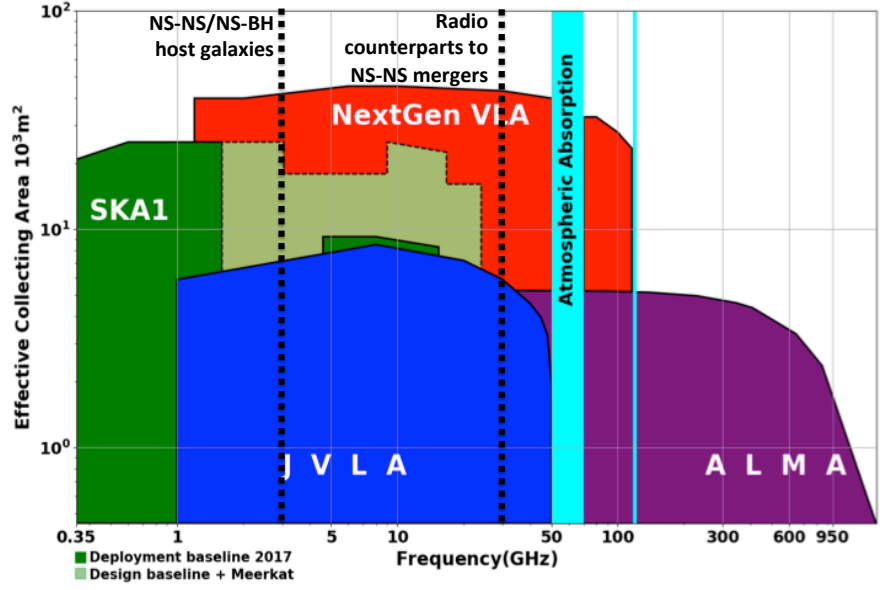}
\vspace{-0.8cm}
\caption{Figure adapted from the ngVLA project summary presentation by E. Murphy. See \url{https://science.nrao.edu/futures/ngvla/documents-publications}. The vertical dotted lines mark the frequencies relevant for this study. At these frequencies, the ngVLA collecting area is superior to that of SKA1 and ALMA. \label{sensitivity}}
\vspace{0.7cm}
\end{center}
\end{figure}

We stress that the ngVLA sensitivity (and resolution) will be optimal for the science topics described in this memo when compared to other radio facilities that we expect to be operational in the ngVLA era. Particularly, as shown in Fig. \ref{sensitivity}, we focus here on two topics that make use of the superior ngVLA (+VLBA) sensitivity and resolution at frequency ranges in between those for which SKA1-MID ($\nu\lesssim 1$\,GHz) and ALMA ($\nu\gtrsim 100$\,GHz) will be the premier radio facilities. 

As current EM follow-up efforts are demonstrating, spectroscopic and multi-wavelength observations of the zoo of optical transients populating the GW localization areas, and of their host galaxies, are going to be key to identifying and removing false positives, as well as breaking degeneracies in the physics behind potential EM counterparts  \citep[e.g.,][]{Abbott2016,Bhalerao2017,Connaughton2016,Copperwheat2016,Cowperthwaite2016,Evans2016,Kasliwal2016,Palliyaguru2016,Savchenko2016,Smartt2016,Bhalerao2017,Corsi2017,Kawai2017,Racusin2017,Stalder2017,Verrecchia2017}. Radio and optical observations, for example, crucially complement each other: while optical emission traces the slower thermally-emitting material, radio probes
the non-thermal fastest-moving ejecta. This was clearly demonstrated in the case of GW170817 \citep{GRB,GW170817Discovery}, the first multi-messenger detection of a NS-NS merger  \citep[e.g.][and references therein]{MMApaper,Coulter2017,Evans2017,Hallinan2017,Kasliwal2017,Troja2017,Valenti2017} where optical/IR observations revealed a kilonova and the r-process nucleosynthesis, while the late-time radio and X-ray emissions probed a completely different component, namely, a fast jet observed off-axis (also powering a short $\gamma$-ray burst; GRB). It is in this context that the value of a PI-driven radio array such as the ngVLA, with a sensitivity and resolution matched to (or encompassing that of) other post-2027 facilities, is best understood. 

In light of the above, hereafter we make the reasonable assumption that by the time the ngVLA becomes operational, the community will have collected statistically significant samples of various type of mergers in the GW domain. We also assume that at least some of the EM counterparts expected to accompany NS-NS and BH-NS mergers will have been discovered, and their host galaxies identified. Although BH-BH mergers may not have EM counterparts, and although individual GW events may have several potential host galaxies within their GW localization volume, in our discussion we assume that the statistical properties of their host galaxies are going to be inferred by analyzing many events simultaneously.

\section{Resolving the host galaxies of compact binary mergers}
\label{sec:2}

GW observations of compact object binaries (NS-NS, BH-NS, or BH-BH) can constrain the properties of the compact object themselves, such as masses, spins, and merger rates. However, understanding their progenitors and formation channels (i.e. how do compact binary systems actually form and evolve) requires the identification of host galaxies, and more generally a detailed study of the merger environment. 

A first constraint on the progenitors of compact binary mergers and their age distribution would be provided by the demographics of their host galaxies, since the distribution of merger timescales impacts the mix of early- and late-type hosts. \textit{Generally speaking, the smaller the merger delay, the stronger the connection with (recent) star formation (SF) rather than with stellar mass alone, and the larger the late-type fraction} \citep[e.g.,][]{Zheng2007}. A young cluster origin would also connect to late-type galaxies \citep[e.g.,][]{Leary2007}, while a globular cluster origin \citep[e.g.,][]{Rodriguez2016} would connect to early-type galaxies. In the case of BH-BH mergers, a connection with AGN, which has been suggested in the context of formation via dynamical interactions, could also be tested via host studies \citep[e.g.][]{Bartos2017}. Moreover, the dependence of the BH-BH merger efficiency on the metallicity may favor dwarf (low-metallicity) hosts and/or merging galaxies with bursts of SF at low metallicity in their assembly history. Overall, a detailed understanding of the galaxy ``response function'' for compact binaries (how often star forming gas of a given metallicity evolves into merging compact binaries) is likely to ultimately require broad-band datasets and the help of stellar population synthesis techniques \citep{Richard2017}.  

Radio observations could contribute significantly to the above studies. Radio can probe the presence of AGNs in the hosts, and could thus help test the proposed formation scenario that connects (heavy) BH-BH mergers to dynamical interactions in dense stellar systems \citep[including globular clusters and AGN, e.g. ][]{Bartos2017}. More generally, radio continuum emission from galaxies traces the SF rate (SFR) during the last $\sim 50-100$\,Myr \citep[e.g.,][]{Murphy2011}, and could thus help track the connection between mergers and recent SFR rate. In fact, due to the age-dust-metallicity degeneracy, the optical/UV emission alone is an unreliable measure of SFR in dusty galaxies. While the FIR is also an extinction free SFR tracer, and could provide more accurate estimates of SFR when combined with optical data, the ease with which SFRs can be measured is also to be taken into account. In this respect, ground-based extinction-free radio observations are advantageous compared to satellite-based FIR observations.  

In addition to the host galaxy SFR (and type) as linked to the merger delay distribution, the study of sub-galactic environments (achievable via \textit{resolved} multi-wavelength studies of the host galaxies) can be used to gain other clues to the progenitors of compact binary mergers. Resolved host galaxy studies of binary mergers with EM counterparts (and thus with accurate localizations) would first of all remove confusion between potential EM counterparts \citep[especially those slowly-evolving radio transients predicted by some models of NS-NS merger ejecta, see e.g., ][]{Hotokezaka2016} and light from the host galaxy. Resolved studies would also allow us to assess the presence or absence of a spatial association with star formation or stellar mass, and potentially track an overall globular cluster distribution for a dominant population of dynamically formed binaries. Finally, the distribution of offsets between the birth and explosion sites, as determined by natal kicks, will be reflected in the fraction of ``hostless'' mergers (those whose projected locations will extend much beyond the visible extent of typical galaxies) and in the measured distribution of offsets of EM counterparts with respect to their host galaxy light \citep[e.g.,][ and references therein]{Belczynski2006,Behroozi2014}.

\begin{figure}
\begin{center}
\includegraphics[width=15cm]{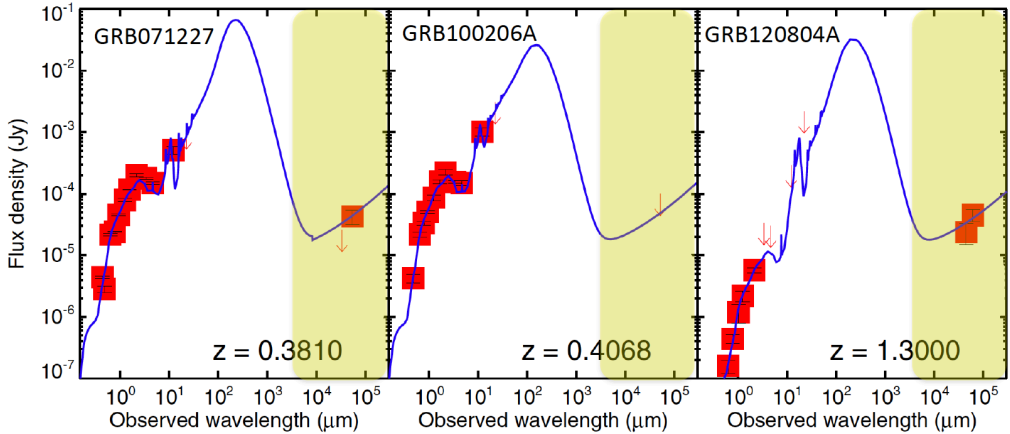}
\vspace{-2.6cm}
\caption{Figure adapted from \citet{Nicuesa2014}. Spectral energy distribution of the three short GRB hosts with (U)LIRG SFRs after correction for Galactic extinction. Red squares are data points, whereas
blue lines denote GRASIL models for the host galaxy emission \citep{Silva1998}. The yellow shaded areas mark the wavelength range of the ngVLA. For GRB\,071227, GRB\,100206A, and GRB\,120804A, data are taken from \citet{Perley2012}, \citet{Berger2013}, and \citet{Nicuesa2014}, respectively. The GRASIL fits assume a radio slope of $-0.75$. For GRB\,071227 and GRB\,120804A, SFR estimated using optical data alone would yield SFR $\approx 10-100\times$ smaller than those obtained including radio data.}
\end{center}
\end{figure}
\subsection{The short GRB host galaxy sample}
To understand the role that ngVLA observations can play in the host galaxy study of compact binary mergers, here we consider the example of short GRBs. Because short GRBs are likely associated with cosmological mergers of compact binaries containing at least one NS and/or stellar-mass BHs, their host galaxies form a good sample to draw from in the context of compact binary mergers detectable by ground-based GW detectors. 

First, we note that radio observations of the host galaxies of GRB\,071227 and GRB\,120804A provided clear evidence for SFRs at least an order of magnitude larger than those derived using optical data alone \citep[Fig. 1;][]{Nicuesa2014}, suggesting either a starburst origin or an AGN contribution. For instance, for the host of GRB\,071227, \citet{Nicuesa2014} find a SFR of $\approx 24$\,M$_{\odot}$/yr (Fig. 1), much in excess to the SFR of $\approx 0.6$\,M$_{\odot}$/yr estimated from the optical data alone (see Table 1). This supports our expectation that optical data alone, due to the age-dust-metallicity degeneracy, cannot reliably determine the host SFR for binary mergers detectable by ground-based GW detectors. The host galaxy NGC4993 of the recently detected binary NS merger GW170817, associated with the short GRB\,170817A, shows a radio SFR of $\approx 0.1$\,M$_{\odot}$/yr, approximately $ 10\times$ higher than the one estimated using the galaxy broad-band photometry, indicative of an AGN dominating the radio emission \citep[][]{Blanchard2017}.

While currently most of the cosmological short GRB host galaxies remain undetected in the radio, a sensitive array like the ngVLA will be able to resolve short GRB-like hosts within the LIGO horizon distance. To show this, in Table 1 we collected the (optical/UV) SFR and optical sizes  of short GRB hosts available in the literature. We use the measured SFR to calculate the expected  host galaxy luminosity at 1.4\,GHz via the \citet{Murphy2011} SFR-to-non-thermal radio emission calibration relation:
\begin{equation}
\left(\frac{\rm SFR_{1.4\rm\,GHz}}{M_{\odot}\rm yr^{-1}}\right)=6.35\times10^{-29}\left(\frac{L_{1.4\rm\,GHz}}{\rm erg\,s^{-1}Hz^{-1}}\right).
\end{equation}
\begin{figure}
\begin{center}
\includegraphics[angle=-90,width=10cm]{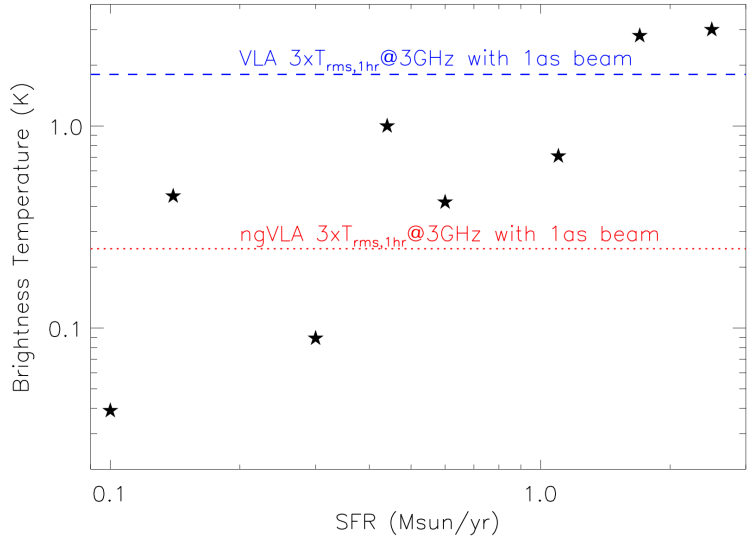}
\vspace{-0.2cm}
\caption{Predictions for the radio brightness temperature of NS-Ns and BH-NS binary merger host galaxies located with the advanced LIGO horizon distance ($z\lesssim 0.1$). These predictions are based on our current knowledge of short GRB hosts (see Table 1). The ngVLA will reach a surface brightness sensitivity of $\approx 0.0823$\,K in 1 hr with $\approx 1''$ resolution at 3 GHz, thus resolving most of these hosts at $z\lesssim 0.1$ (red dotted line). We note that resolved studies of most short GRB-like host galaxies within the advanced LIGO horizon are inaccessible to the Jansky VLA for reasonable integration times (blue dashed line).}
\end{center}
\end{figure}
To estimate the corresponding luminosity at the lowest ngVLA frequency of $\approx 3$\,GHz, we extrapolate from the 1.4\,GHz luminosity using a spectral index of $\approx -0.75$ (as appropriate for non-thermal galaxy emission). The results are reported in Table 1, together with the corresponding expected radio flux densities at $z=0.1$ (the advanced LIGO horizon distance). We then use the effective galaxy radius $r_e$ from optical/IR observations as FWHM size of the galaxy at radio wavelengths. This is justified by the fact that radio emission from local galaxies is typically observed to be more concentrated than in the optical. Specifically, using the \citet{Condon2002} sample, we estimate an average ratio of $\approx 0.5$ between the FWHM of the radio galaxies in the NVSS catalog (FWHM$\approx 45''$) and the size of their blue diameter. The smaller concentration of the radio emission in star forming galaxies has also been confirmed in the GOODS-N Jansky VLA sample by comparing effective radii of the 10\,GHz emission with effective radii derived from HST/WFC3 observations \citep{Murphy2017}. With the expected radio flux and expected radio size in hand, we thus calculate the expected brightness temperature as:
\begin{equation}
T=0.32\times10^{23}\frac{\lambda^2}{\theta^2}S,
\end{equation}
where $\lambda$ is the observed wavelength, $\theta$ is the FWHM radio size, and $S$ is the flux density at the considered wavelength ($\lambda \approx 10$\,cm for observations at 3\,GHz). Our results are reported in the last column of Table 1, and also plotted in Fig. 2. The ngVLA will reach a surface brightness sensitivity of $\approx 0.0823$\,K ($\approx 0.18$\,K) in 1 hr with $\approx 1''$ ($\approx 0.1''$) resolution at 3 GHz, thus resolving most of these hosts at $z\lesssim 0.1$. We stress that the current VLA in its A configuration, would offer a comparable resolution of $\approx 1''$ at 3\,GHz, but with a worse surface brightness sensitivity in 1\,hr of $\gtrsim 0.6$\,K. Only two of the host galaxies in Table 1 have a brightness temperature larger than $\approx 3\times 0.6\,{\rm K}=1.8$\,K, thus \textit{resolved studies of most short GRB-like host galaxies within the advanced LIGO horizon would be inaccessible to the Jansky VLA (for reasonable integration times, see the blue horizontal line in Fig. 2) and require a more sensitive array like the  ngVLA}.

\begin{table}
\begin{center}
\caption{Host galaxy properties of short GRBs. In the left hand side, we collect literature data of short GRB host galaxy SFR and effective optical radii. In the right hand side, we summarize expectations for the radio FWHM, 3\,GHz luminosity, 3\,GHz flux density, and brightness temperatures of BH-NS/NS-NS host galaxies with similar SFR and physical sizes, located within the advanced LIGO horizon of $z=0.1$  (see text for details). The ngVLA will reach a surface brightness sensitivity of $\approx 0.0823$\,K ($\approx 0.18$\,K) in 1 hr with $\approx 1''$ ($\approx 0.1''$) resolution at 3 GHz, thus resolving most of these hosts at $z\lesssim 0.1$.}
\begin{tabular}{llclc|cccc}
\hline
GRB  &   $z$ &  SFR  & $r_{e, z}$ & $r_e$ & $FWHM_{z=0.1}$ & $L_{3\,\rm GHz}$ &$F_{3\,{\rm GHz}, z=0.1}$ & T\\
          &          & $(M_{\odot}$/yr) & ($''$) &  (kpc) & ($''$) & (erg\,s$^{-1}$\,Hz$^{-1}$) &($\mu$\,Jy) & (K)\\
\hline
061201   & 0.111 & 0.14  &1.09 & 2.2 & 1.2 & $1.2\times10^{27}$ & 4.8 & 0.45\\
070429B  & 0.902 & 1.1 & 0.65 & 5.1 & 2.7 & $9.7\times10^{27}$ & 38 & 0.71\\
070714B  & 0.922  & 0.44 & 0.34 & 2.7 & 1.4 & $3.9\times10^{27}$  & 15 & 1.0\\
070724A  & 0.457  & 2.5 & 0.63 & 3.7 & 2.0 & $2.2\times10^{28}$ & 87 & 3.0\\
071227   & 0.381 & 0.6 & 0.91 & 4.8 & 2.6 &$5.3\times10^{27}$ & 21 & 0.42\\
090510   & 0.903 & 0.3  & 0.93& 7.3 & 3.9 &$2.7\times10^{27}$ & 10 & 0.089 \\
090515   & 0.403  & 0.1  & 1.19 & 6.5 & 3.5 & $8.9\times10^{26}$  & 3.5 & 0.039\\
130603B & 0.356 & 1.7 & 0.62 & 3.1 & 1.7 & $1.5\times10^{28}$ & 59& 2.8 \\
\hline
\end{tabular}
\end{center}
\end{table}

\section{Direct size measurement of relativistic ejecta from NS-NS and BH-NS mergers}
\label{sec:3}
NS-NS and BH-NS mergers are thought to produce, in addition to GW in-spiral signals, EM (and, specifically, synchrotron radio) emission associated with outflows with significant kinetic energies. This prediction has been proven accurate by GW170817, a NS-NS merger associated with emission of EM radiation at all wavelengths \citep[e.g.][and references therein]{MMApaper,Evans2017,Hallinan2017,Kasliwal2017,Troja2017,Valenti2017}. In what follows, we show that the ngVLA with its superior sensitivity could probe, for the first time directly, the early-time dynamics of relativistic ejecta from NS-NS and BH-NS mergers located within the distance horizon of ground-based GW detectors. Resolving the radio remnant of a relativistic ejecta requires VLBI techniques. So far, given the large cosmological distances of most GRBs detected via purely EM observations, this has been attempted only for the brightest \textit{long} GRBs ($\gamma$-ray transients associated with massive star collapse), and limited to long timescales after the explosion (so as to give time to the ejecta to expand). \textit{Direct imaging of the relativistic ejecta associated with BH-NS or NS-NS producing short GRBs, which generally have lower luminosities and dimmer radio afterglows compared to long GRBs, has not been within the reach of current radio arrays, and represents the only direct way to map the speed distribution of merger ejecta and test the predictions of dynamic ejecta models}. 

Broadly speaking, three major classes of radio counterparts are thought to exist \citep[e.g.,][]{Hotokezaka2016,Lazzati2017}: (i) Counterparts associated with sub-relativistic merger ejecta producing radio remnants on timescales of a few years; (ii)  ultra-relativistic jets that produce short GRBs and (on-axis) radio afterglows in the direction of the jet (evolving on  timescales of a few days), and/or orphan-radio afterglows extending over wider angles (evolving on timescales of weeks); (iii) mildly-relativistic ``cocoons" whose emission may be more isotropic and whose brightness would depend on the amount of energy imparted to it by the jet. Here, we consider scenario (ii) and focus on the case in which the observer is located on-axis with an ultra-relativistic jet launched during the merger. Other emission processes and geometries may be possible, but this example gives a good order of magnitude estimate of what is possible with the VLBI capabilities of ngVLA. In this case, detailed calculations for the predicted surface brightness of the radio counterpart exist \citep[e.g.,][]{Granot1999}. According to these predictions, the image of a short GRB afterglow would to be ring-like: brighter near the edge and dimmer near the center, with the contrast between the edge and the center of the image increasing (and the width of the ring decreasing) at higher frequencies for a given observing time \citep[see Fig. 4, left, and ][]{Granot1999}. 

\begin{figure}
\begin{center}
\includegraphics[width=5cm]{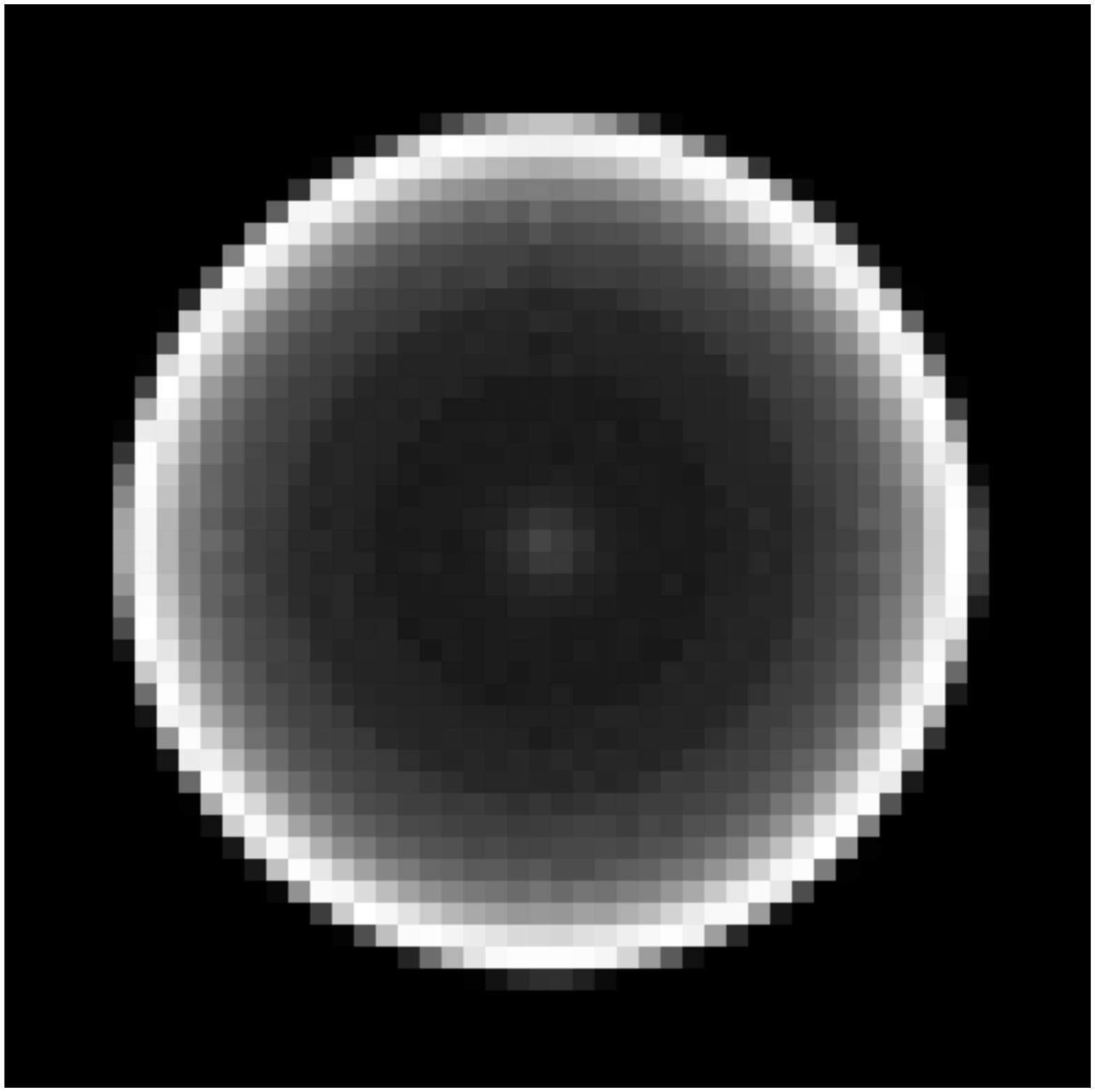}
\hbox{\hspace{1.3cm}\includegraphics[height=7cm]{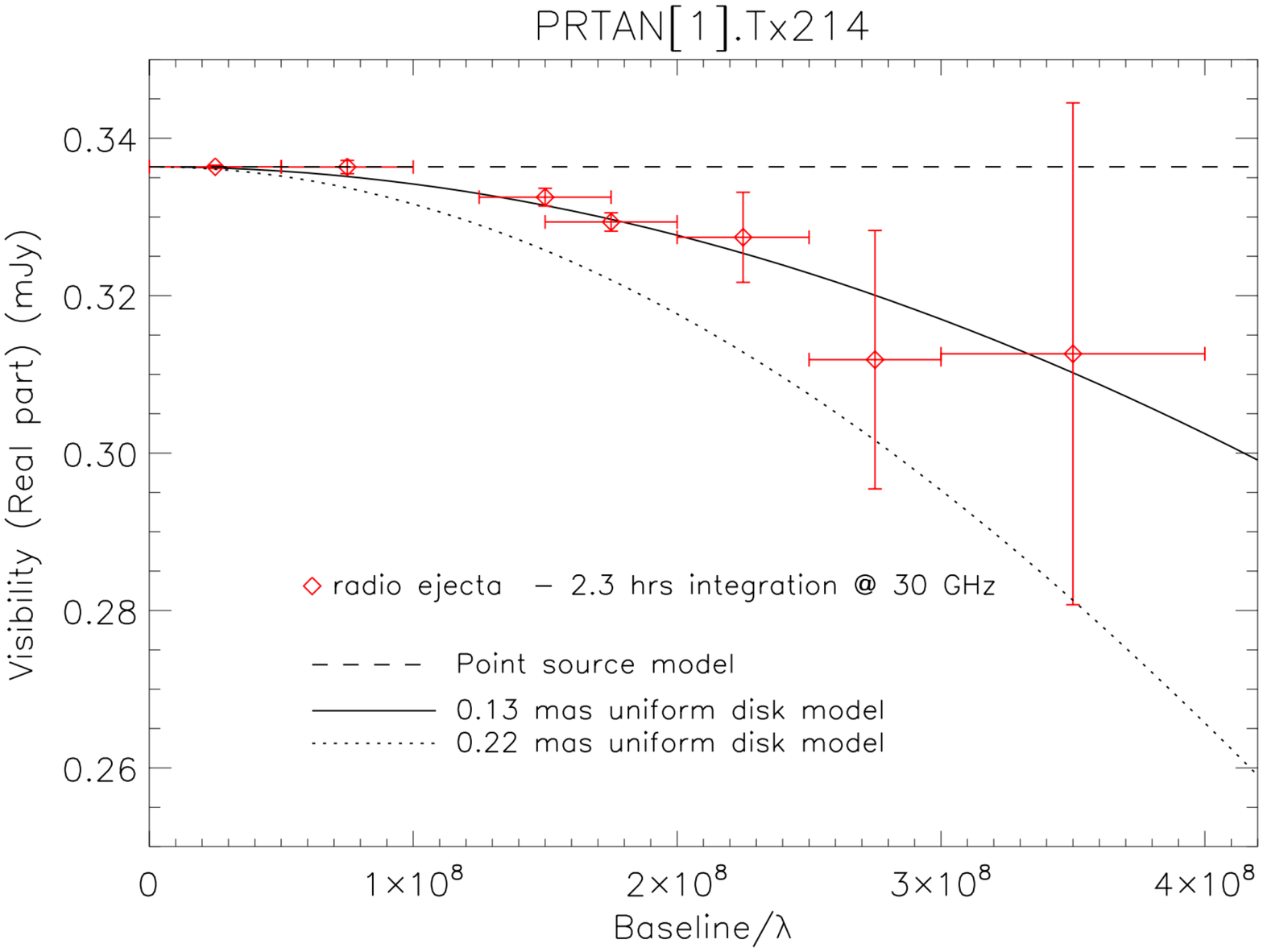}
}
\vspace{-0.2cm}
\caption{Left: Figure adapted from \citet{Granot1999}. We show the expected image of a GRB afterglow for $\phi=\frac{\nu}{\nu_m}\sim10$ where $\nu_m$ is the time-evolving characteristic synchrotron frequency of electrons located just behind the shock front on the line of sight (see text). This image has been created by smoothing the corresponding $\phi=10$ image presented in \citet{Granot1999}. Larger values of $\phi$ represent a higher observing frequency or a later observing time (since $\nu_m\propto t^{-3/2}$). The image is brighter near the edge and dimmer near the center, and the contrast is higher at larger values of $\phi$. Right: Simulation of a 2.3\,hr-long ngVLA observation of the radio remnant of a relativistic ejecta from a BH-NS/NS-NS merger at $d_{L}\approx 150$\,Mpc, with realistic parameter values of $E_{iso}\approx 1.8\times10^{50}$\,erg, $\epsilon_e=0.1$, $\epsilon_B=0.01$, $n=0.01$\,cm$^{-3}$. We have used the model image shown in the left panel, assumed an ngVLA configuration as shown in Fig. 5 (left panel), and added noise level compatible with current estimates of expected  ngVLA performance at 30\,GHz (see Fig. 5, right panel).}
\end{center}
\end{figure}

\begin{figure}
\begin{center}
\hbox{
\includegraphics[height=7cm]{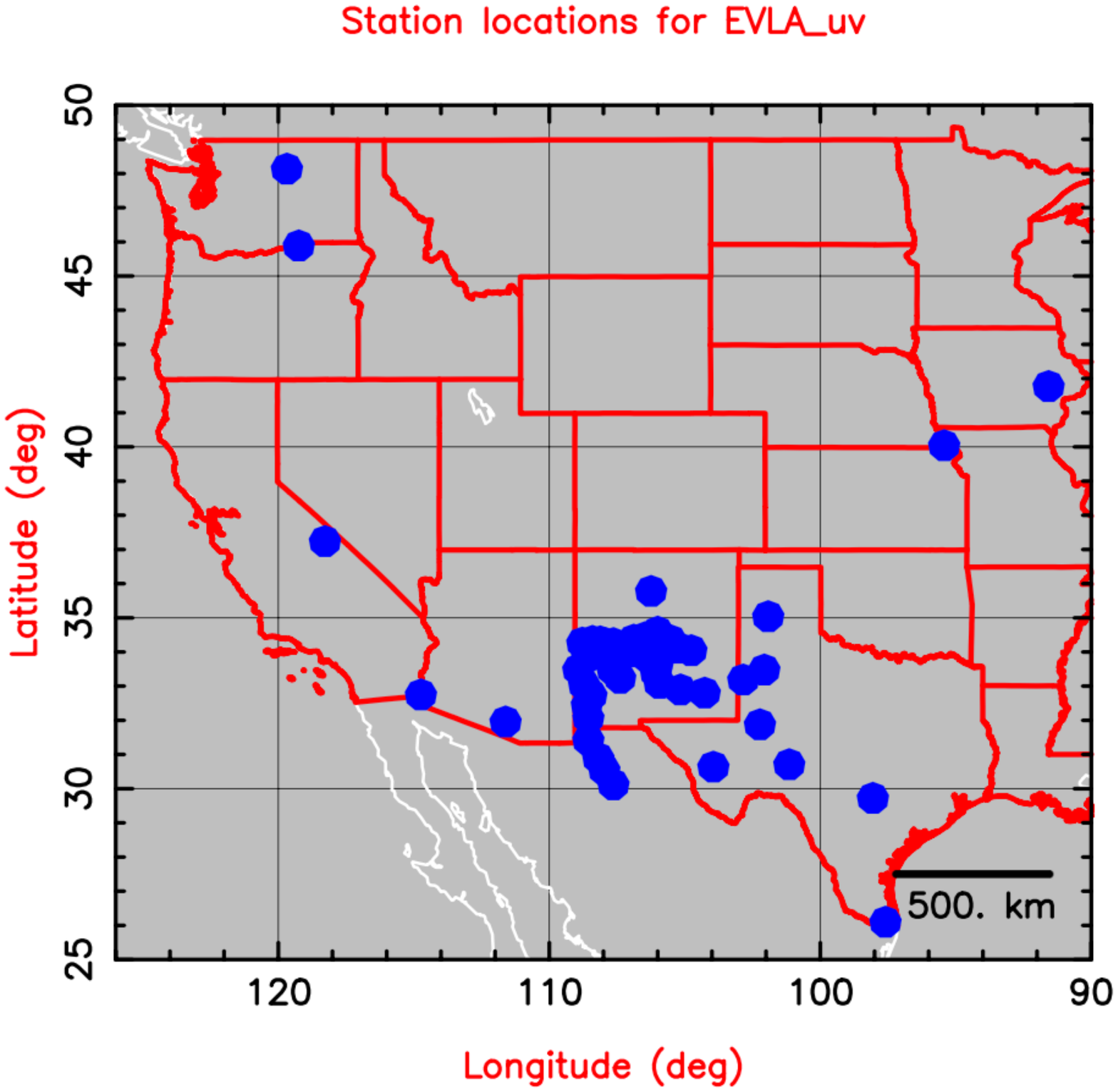}
\includegraphics[height=7cm]{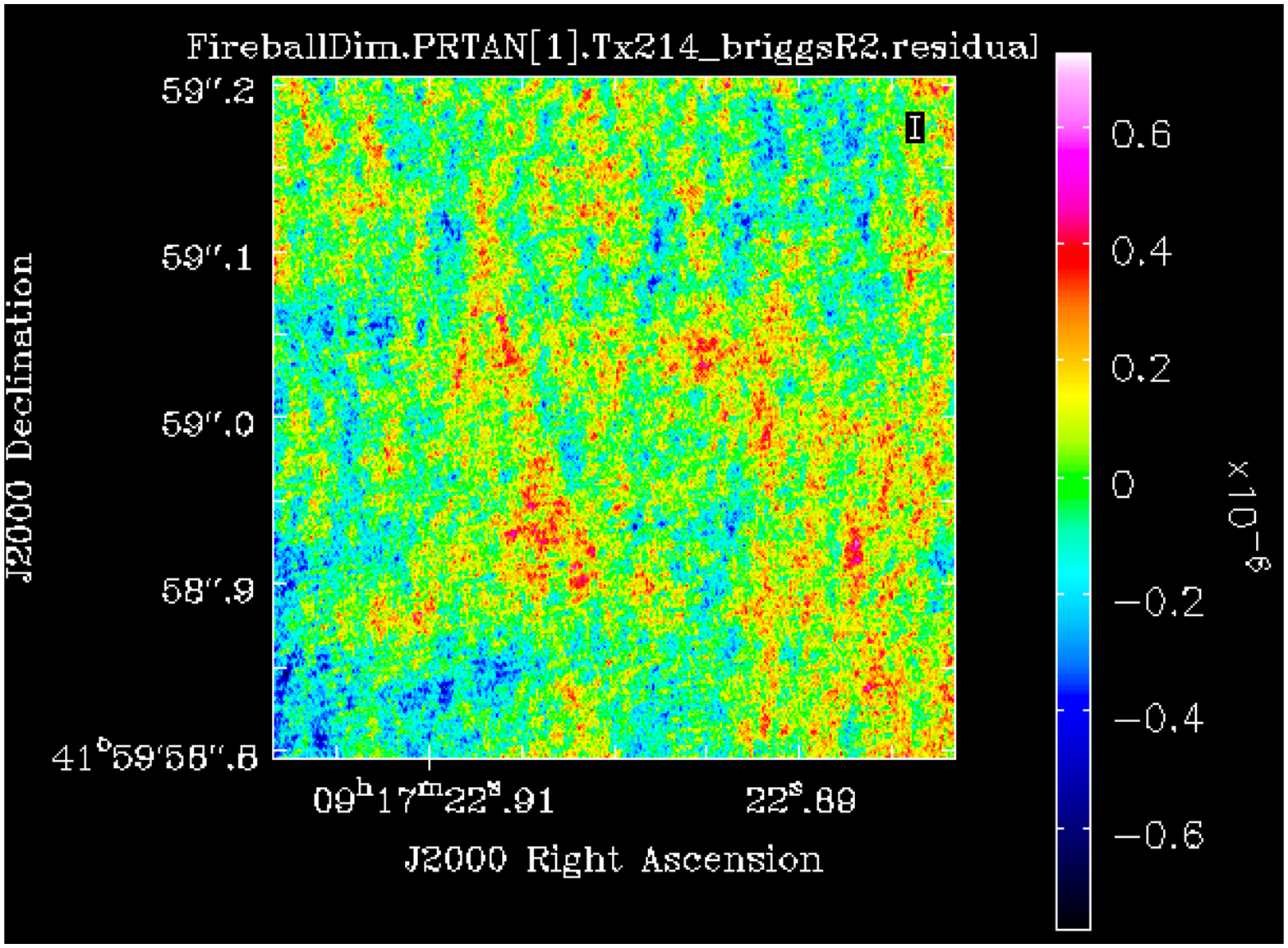}}
\caption{Left: ngVLA antennas configuration used in the simulation shown in Fig. 4, where VLBA stations have also been added. Right: Simulated ngVLA residual noise image in 2.3\,hr integration time at 30\,GHz (rms of $\approx 0.15\,\mu$Jy) with natural weighting (robust=2) as expected for the ngVLA. }
\end{center}
\end{figure}

Following \citet{Granot1999}, we write the characteristic synchrotron frequency of electrons located just behind the relativistic shock on the line of sight as:
\begin{equation}
\nu_m\approx 1.3\times10^{9}\,{\rm Hz}\sqrt{1+z}\left(\frac{\epsilon_B}{0.01}\right)^{1/2}\left(\frac{\epsilon_e}{0.1}\right)^{2}\left(\frac{E_{\rm iso}}{3\times10^{50}\,{\rm erg}}\right)^{1/2}\left(\frac{t}{7\,{\rm d}}\right)^{-3/2},
\end{equation}
where $\epsilon_e$ and $\epsilon_B$ denote the fraction of energy going into accelerating relativistic electrons and amplifying magnetic fields (with typical values of $\epsilon_e\approx 0.1$ and $\epsilon_B\approx 0.01$), and $E_{iso}$ is the isotropic-equivalent kinetic energy of the ejecta. We note that $E_{iso}\approx 1.8\times10^{50}$\,erg corresponds to a true jet energy of $E_{j}\approx 2.5\times10^{49}(\theta_j/30\,\deg)^2$\,erg, where $\theta_j$ is the jet half-opening angle. The synchrotron emission spectrum peaks at a frequency which, for a typical value of the power-law index of the electron energy distribution of $p\approx 2.5$, is $\approx 1.88\times \nu_m$ \citep{Granot1999}:
\begin{equation}
\nu_{p}\approx 2.4\times10^{9}\,{\rm Hz}\sqrt{1+z}\left(\frac{\epsilon_B}{0.01}\right)^{1/2}\left(\frac{\epsilon_e}{0.1}\right)^{2}\left(\frac{E_{\rm iso}}{1.8\times10^{50}\,{\rm erg}}\right)^{1/2}\left(\frac{t}{7\,{\rm d}}\right)^{-3/2}.
\end{equation}
For a merger ejecta located within the advanced LIGO horizon distance ($z\lesssim 0.1$), we can approximate the total flux received by an on-axis observer at  $\nu\approx 30$\,GHz and $t\approx 7$\,d as:
\begin{equation}
F_{\nu}\approx 0.33\,{\rm mJy}\,(1+z)^{11/8}\left(\frac{\epsilon_B}{0.01}\right)^{7/8}\left(\frac{\epsilon_e}{0.1}\right)^{3/2}\left(\frac{E_{\rm iso}}{1.8\times10^{50}\,{\rm erg}}\right)^{11/8}\left(\frac{n}{0.01}\right)^{1/2}\left(\frac{d_L}{150\,{\rm Mpc}}\right)^{-2}\left(\frac{\nu}{30\,{\rm GHz}}\right)^{-3/4}\left(\frac{t}{7\,{\rm d}}\right)^{-9/8},
\end{equation}
where $n$ denotes the density of the ISM in units of cm$^{-3}$. The above flux is distributed as shown in Figure 3 (left panel), with an observed outer radius of the ring equal to  \citep{Granot1999}:
\begin{equation}
R_{\perp,\rm max}\approx 1.4\times10^{17}\,{\rm cm}\left(\frac{E_{\rm iso}}{1.8\times10^{50}\,{\rm erg}}\right)^{1/8}\left(\frac{n}{0.01}\right)^{-1/8}\left(\frac{t}{(1+z)7\,{\rm d}}\right)^{5/8}.
\end{equation}
Thus, for a source located at $ 150$\,Mpc, the outer angular diameter of the relativistic ejecta image can be estimated as:
\begin{equation}
D_{\rm max, 150\,Mpc}\approx 0.13\,{\rm mas}\left(\frac{E_{\rm iso}}{1.8\times10^{50}\,{\rm erg}}\right)^{1/8}\left(\frac{n}{0.01}\right)^{-1/8}\left(\frac{t}{7\,{\rm d}}\right)^{5/8}.
\end{equation}
We note that the above expressions for the radio frequencies, fluxes, and size of the radio remnant are valid for all time $t\lesssim t_j$ where \citep{Granot2005}:
\begin{equation}
t_j\approx 26\,{\rm d}\,(1+z)\left(\frac{E_j}{2.5\times10^{49}\,{\rm erg}}\right)^{1/3}\left(\frac{n}{0.01}\right)^{-1/3}\left(\frac{\theta_j}{30\deg}\right)^2.
\end{equation} 

In Figure 4, right panel, we show a simulation of a 2.3\,hr-long ngVLA observation of the radio remnant of a relativistic ejecta from a BH-NS/NS-NS merger at $d_{L}\approx 150$\,Mpc, with realistic parameter values of $E_{iso}\approx 1.8\times10^{50}$\,erg, $\epsilon_e=0.1$, $\epsilon_B=0.01$, $n=0.01$\,cm$^{-3}$. We have used the model image shown in the left panel of Fig. 4, assumed an ngVLA configuration as shown in Fig. 5 (left panel) with the addition of VLBI stations with baselines up to $\approx 3500$\,km, and added noise level compatible with current estimates of expected  ngVLA performance at 30\,GHz (see Fig. 5, right panel). As evident from this Fig. 4 (right panel), with the ngVLA we will be able to measure sizes as small as $\approx 0.13$\,mas at flux levels of order $\approx 0.33$\,mJy. We can compare this result with previous VLBI size measurements performed on the radio afterglows of long duration GRBs. Particularly, \citet{Taylor2004} reached similar size constraints in comparably long observations of GRB\,030329 ($z\approx 0.17$) with VLBA plus other stations. However, GRB\,030329 had fluxes $\gtrsim 3$\,mJy in the $8-22$\,GHz range at the times at which these size constraints were obtained. \textit{Thus, our simulation shows that the ngVLA would enable size measurements comparable to those that have been achieved in the past for long GRBs (which are associated to the deaths of massive stars) using the VLA plus VLBI stations (with similar integration times), but for radio remnants $\approx 10\times$ dimmer, which is key for probing NS-NS and BH-NS mergers - the golden multi-messenger sources whose radio size measurements have thus far remained outside the reach of current radio arrays}.

\section{Summary and conclusion}
\label{sec:4}
We have investigated two new scientific opportunities that would emerge in time-domain astrophysics if a facility like the ngVLA were to work in tandem with ground-based GW detectors:
\begin{enumerate}
\item Unraveling the physics behind the progenitors of BH-BH, NS-NS, and BH-NS mergers via host galaxy studies at radio wavelengths;
\item Enabling direct size measurements (and thus direct dynamics constraints) of relativistic ejecta from NS-NS and/or BH-NS mergers (with ngVLA plus VLBI stations).
\end{enumerate}
We have shown that:
\begin{enumerate}
 \item The ngVLA, thanks to its improved surface brightness sensitivity at lower frequencies, can enable resolved studies of NS-NS and/or BH-NS mergers host galaxies within the advanced LIGO horizon distance. Based on current knowledge drawn from the short GRB host galaxy sample, these galaxies would otherwise remain largely inaccessible to the Jansky VLA (for realistic integration times).
 \item The ngVLA+VLBI, thanks to its superior high frequency ($\nu\approx 30$\,GHz) sensitivity at the longest baselines, can enable direct size measurements of relativistic ejecta from NS-NS and/or BH-NS mergers to a level ($\sim 0.1$\,mas) comparable to what has been achieved in the past for long GRBs (using the VLA plus VLBI stations with similar integration times) but for radio remnants $\approx 10\times$ dimmer, which is key for probing directly the currently unexplored dynamics of relativistic ejecta from binary NS of BH-NS mergers.
\end{enumerate}
We finally stress that having the ngVLA be operated as a PI-driven array capable of fast response to GW triggers is fundamental to enabling the scientific results described in this memo.\\

\acknowledgments
A.C. thanks Chris Carilli and Remy Indebetouw for many discussions about the ngVLA capabilities, and for prompt help with simulations of ngVLA observations within CASA. In fact, this study would not have been possible without the help of Chris Carilli and Remy Indebetouw. The National Radio Astronomy Observatory is a facility of the National Science Foundation operated under cooperative agreement by Associated Universities, Inc.

\bibliographystyle{aasjournal}
\bibliography{Corsi_Memo}

\end{document}